# Influence of π – iodide intermolecular interactions on electronic properties of tin(IV) iodide semiconducting complexes

*Ewelina Wlaźlak,[a,b] Wojciech Macyk,[a] Wojciech Nitek,[a] Konrad Szaciłowski[b]\**

[a]Faculty of Chemistry, Jagiellonian University, ul. R. Ingardena 3, 30-060 Kraków, Poland

[b]AGH University of Science and Technology, Academic Centre for Materials and Nanotechnology, al. A. Mickiewicza 30, 30-059 Kraków, Poland

*corresponding author: szacilow@agh.edu.pl

ABSTRACT: Coordination compounds with tin centre surrounded by both organic and inorganic ligands ([SnI$_4${(C$_6$H$_5$)$_3$PO}$_2$], [SnI$_4${(C$_6$H$_5$)$_2$SO}$_2$] and [SnI$_4$(C$_5$H$_5$NO)$_2$]), acting as molecular semiconductors are in the spotlight of this article. This is a new class of hybrid semiconducting materials where optoelectronic properties of inorganic core ($SnI_4$) were tuned by organic ligands. The valence band is located at inorganic part of the molecule while the conduction band is made of carbon-based orbitals. This suggests a great importance of hydrogen bonds where iodine atoms play the role of an acceptor. Weak intermolecular interactions between iodine atoms and aromatic rings are essential in a band structure formation. These materials form orange-red crystals soluble in most of organic solvents. Their semiconducting properties are addressed experimentally via photovoltage measurements, as well as theoretically, using DFT and semiempirical approaches.

**Introduction**

Further development of photovoltaics and optoelectronics requires easily processable materials with tunable electronic and optical properties. Their application may result in devices with higher efficiencies and better parameters. So far, hybrid semiconducting materials in which molecules of metal compounds interact with each other only by weak forces, have been studied only scarcely. This group of materials has interesting electronic properties, therefore more detailed studies are required. Semiconductors can be divided into two main groups: inorganic and organic materials (Figure 1). These classes of materials differ not only at the chemical level, but more importantly, they possess completely different electronic properties due to the differences in their atomic and electronic structures. In inorganic semiconductors atoms are held together by strong covalent and ionic interactions whereas weak van der Waals and hydrogen interactions determine properties of organic semiconductors. Differences in intramolecular interactions are the most important factor that determines delocalization of electrons and their wave functions over the crystal lattice and thus the electronic properties of a semiconducting material.

Strong bonds (covalent, ionic) induce formation of band structure where allowed energy states form continuous conduction and valence bands, with a broad bandwidth (inorganic semiconductors, like Si or ZnS). On the other hand, weak van der Waals and hydrogen bonds dominate in organic crystals. The strength of the intermolecular interactions is reflected in their band structure and density of states.[1] The weaker intermolecular interactions are, the thinner and the less continuous bands are formed.[2] Organic semiconductors (polymers and small molecules) have in common a π-electron system formed by the $p_z$ orbitals of $sp^2$ hybridized carbon atoms.[3] In organic semiconductors weak van der Waals and hydrogen interactions govern the formation of crystal structures. In hybrid materials both scenarios are possible. Covalently bonded metal halides and charged layers held together by ionic interactions make perovskites more similar to inorganic semiconductors, in which the layered structure provides a suitable arrangement for a high charge carrier mobility.[4] Hybrid materials combine properties of organic and inorganic semiconductors. Perovskites have the corner sharing metal halide octahedra that form an anionic lattice and organic cations form a network for the inorganic structure. Observed properties in this group of materials have the origin in the connectivity of the inorganic sublattice– organic cations are used to tune the electronic properties by structural templating.[5, 6] There are also materials which cannot be classified as perovskites, but are based on similar building blocks.[7-9] For example, the anionic metal halide octahedra lattice was found in a hybrid material where negative charge of inorganic sheets is compensated by a non-innocent tropylium cation, involved in a charge transfer interaction within the lattice of ($C_7H_7PbI_3$).[10] In this compound the electrostatic interactions between organic and inorganic components held the structure together.

In neutral $SnI_4L_2$ complexes described in this paper, in contrary to the inorganic and the majority of hybrid semiconductors, the weak interactions are responsible for formation of the band structure. These materials are hybrid semiconductors with tin(IV) iodide complexed by two organic ligands. Organic and inorganic molecules are connected by coordination bonds between oxygen and tin atoms unlike the other hybrid materials, molecules of these materials are held together by weak van der Waals interactions and hydrogen bonds.

The crystal structure of tin tetraiodide was reported in literature for the first time in 1923.[11] Since then the topic of $SnI_4$ properties has been discussed in several papers, mostly during 60-80's. $[SnI_4\{(C_6H_5)_3PO\}_2]$ and $[SnI_4\{(C_6H_5)_2SO\}_2]$ were mentioned in literature but mostly in the context of the Mössbauer spectroscopy, whereas their electronic properties were not investigated.[12-14] In this paper $[SnI_4\{(C_6H_5)_3PO\}_2]$, $[SnI_4\{(C_6H_5)_2SO\}_2]$ and $[SnI_4(C_5H_5NO)_2]$ complexes were studied. The influence of different ligands on electronic properties and correlation between intermolecular interactions and the electronic structure have been described on the basis of theoretical calculations, experimental studies and a crystal structure analysis. A particular attention was paid to the CH···I interaction which influences properties of these complexes.

**Experimental**

*Syntheses*

*$SnI_4$*: Tin powder (1 g) and iodine (3 g) were heated at 40 °C in anhydrous dichloromethane (100 mL), under reflux until the reaction mixture turned dark orange. The solution was then filtered, concentrated to 50 mL and cooled in an ice bath. Precipitated orange solid was filtered to remove unreacted tin, dried and recrystallized from chloroform, yielding orange crystals of $SnI_4$ (2.7 g). Yield: 51%.

*[SnI$_4${(C$_6$H$_5$)$_3$PO}$_2$]:* Triphenylphosphine oxide (2.1 mmol) and tin tetraiodide (1 mmol) were dissolved in 15 mL of anhydrous dichloromethane and stirred for 1 h at room temperature. The solution was concentrated to 7 mL and cooled. The dark red participate was filtered off, washed with anhydrous dichloromethane and dried. The final product was recrystallized from mixture of cyclohexane and isobutanol. Elemental analysis showed: C- 36.48%, H- 2.59%. Calculated for C$_{36}$H$_{30}$P$_2$O$_2$SnI$_4$: C- 36.55%, H- 2.55%. Yield: 60 %.

*[SnI$_4${(C$_6$H$_5$)$_2$SO}$_2$]*: Diphenyl sulphoxide (2.1 mmol) and tin tetraiodide (1 mmol) were dissolved in 15 mL of anhydrous dichloromethane and stirred for 1 h at room temperature. Subsequently dichloromethane was evaporated under reduced pressure. As the result a dark red solid was obtained. The final product was isolated by crystallization from acetonitrile as fine, red crystals with orange shine. Elemental analysis: C- 27.91%, H- 1.96%, S- 6.21%. Calculated for C$_{24}$H$_{20}$S$_2$O$_2$SnI$_4$: C- 27.96%, H- 1.96%, S- 6.22%. Yield: 27%.

*[SnI$_4$(C$_5$H$_5$NO)$_2$]*: Pyridine-N-oxide (2.1 mmol) and tin tetraiodide (1 mmol) were dissolved in 15 mL of anhydrous dichloromethane, and stirred for 1 h at room temperature. The solution was concentrated to 7 mL and cooled. The red-orange participate was filtered off, washed with anhydrous dichloromethane and dried. The final product was recrystallized from acetonitrile. Elemental analysis: C- 14.74%, H- 1.23%, N- 3.51%. Calculated for C$_{10}$H$_{10}$N$_2$O$_2$SnI$_4$: C- 14.71%, H- 1.23%, N- 3.43%. Yield: 18%.

*Caution! These syntheses are potentially hazardous due to toxic, irritating and lachrymatory properties of iodine. Appropriate means of personal protection should be applied. All wastes should be properly disposed due to high toxicity towards aquatic organisms.*

*X-Ray crystallography*

Data collection and reduction: Diffraction data for single crystal were collected at 120 K using the Oxford Diffraction SuperNova four circle diffractometer, equipped with the Mo (0.71069 Å) Kα radiation source, graphite monochromator and Oxford CryoJet system for measurements at low temperature. Cell refinement and data reduction were performed using the firmware.[15] Structure solution and refinement: Positions of all of non-hydrogen atoms were determined by direct methods using SIR-97.[16] All non-hydrogen atoms were refined anisotropically using weighted full-matrix least-squares on F2. Refinement and further calculations were carried out using SHELXL-97.[17] Graphics were created by MERCURY.[18] Whole used software is a component of the WINGX.[19] Hydrogen atoms treatment: All hydrogen atoms joined to carbon atoms were positioned with an idealized geometry and refined using a riding model with Uiso(H) fixed at 1.2 Ueq of C.

*Computational details*

Optimized molecular structures of the studied complexes were obtained from experimental crystal structure. Firstly, structures were optimized in vacuum (UFF) using the Avogadro software package,[20] then optimized again with the Polarizable Continuum Model (PCM, acetonitrile) using the Gaussian 09 software package[21] at the HF level of theory theory with 6-31+G(d,p)[22] basis set for C, O and P atoms and DGDZVP[23] basis set for Sn and I atoms. Mixed basis sets were used to shorten the initial calculations time. Based on optimized structures of complexes, UV-Vis simulations and charge decomposition analysis were performed.

Simulation of UV-Vis absorption spectra was performed with Time Dependent Density functional theory (TDDFT) using Gaussian 09, PCM solvation model (acetonitrile), DGDZVP[23] basis set for all atoms and RB3lYP[24, 25] (hybrid density functional model). GaussSum software[26] was used to calculate the lowest 50 transitions. Obtained results were used to prepare electrostatic visualizations of frontier orbitals of single molecules with GaussView 5 software.[27] From this calculations also frontier orbitals diagrams and theoretical band gap values were obtained. Visualizations of frontier orbitals of dimers were prepared for two neighboring molecules in crystal, without any optimization, by applying TDDFT calculations, Gaussian 09 and GaussView 5 software. Density of states distributions were calculated for clusters consisting of 12 molecules, using MOPAC2012[28] with the PM7 method and post-processed with the AOMix package.[29, 30] Charge decomposition analysis was performed using AOMix program and population analysis performed by the Gaussian 09 program DFT theory with the DGDZVP basis set.

*Kelvin probe measurements*

Surface photovoltage (SPV) measurements were performed using Kelvin probe-based surface photovoltage spectrometer (Instytut Fotonowy, Poland, and Besocke Delta Phi, Germany) using 150 W xenon arc lamp with a monochromator. Work function (WF) and Fermi level (FL) were measured with Kelvin probe. Samples before measurements were deposited on ITO foil.

*Resistivity measurements*

Room temperature resistivity measurements were performed with Keithley 4200 SCS. Thin layers of dissolved compound were spin coated on glass substrate and measured with four point probe technique. Film thicknesses were measured using Bruker's DektakXT ™ Stylus Profiler.

*UV-Vis spectroscopy, $^1$H NMR and elemental analysis.*

Diffuse reflectance spectra were recorded on Lambda 950 (Perkin Elmer, USA) spectrophotometer. Kubelka-Munk function was used to convert the diffuse reflectance spectra into absorption spectra. Absorption spectra were recorded with HP 8453. $^1$H NMR spectra were recorded with Bruker Avance III 600 MHz. Elemental analyses were taken with elemental micro-analyzer CHNS Vario Micro Cube.

**Results and discussion**

*Crystal structure and interactions*

Crystal structures of [SnI$_4${(C$_6$H$_5$)$_3$PO}$_2$], [SnI$_4${(C$_6$H$_5$)$_2$SO}$_2$] and [SnI$_4$(C$_5$H$_5$NO)$_2$] obtained by a single crystal x-ray diffraction enables a detailed analysis of inter– and intramolecular interactions. As a result of steric intramolecular interactions, SnI$_4$ forms *cis* adducts with triphenylphosphine oxide, diphenyl sulfoxide and pyridine *n*-oxide. Crystal structures of these complexes are present in Figures 2–4. Hydrogen bonding with iodide acceptors (Table 1) seem to play an important role in this group of complexes. The distance (*d*) and angle (*α*) of H⋯I-Sn interaction have been used to determine whether they are consistent with a hydrogen bond description (preferable *d* < 3.14 Å, 90°< *α* < 180°, see Brammer et al.).[31] Stacking distances and selected interatomic distances of iodide are collected in Table 1.

*Crystal structure of the [SnI$_4${(C$_6$H$_5$)$_3$PO}$_2$] complex.*

[SnI$_4${(C$_6$H$_5$)$_3$PO}$_2$] (Figure 2a) crystalizes in a monoclinic space group P2$_1$. There are one Sn$^{4+}$ cation, four coordinated I$^-$ anions and two coordinated triphenylphosphine oxide ligands in the asymmetric unit. The Sn cation is in a distorted octahedral configuration where it is coordinated by four I$^-$ anions as well as by O1 and O2 atoms from triphenylphosphine oxide molecules. The Sn–O bond distances (2.142 and 2.128 Å) are the shortest among other complexes, with O1–Sn–O2 angle of 78.25°. Unlike perovskite, the presented structure lacks strong interactions within the crystal lattice. The role of stacking in semiconductors is well known (in molecular semiconductors the coupling between conjugated systems allows charge to travel through the crystal lattice), but there is only a little information about the role of a weak I⋯H hydrogen bond on the band structure formation. These weak interactions are often found in crystal structures. In [SnI$_4${(C$_6$H$_5$)$_3$PO}$_2$] complex two intermolecular I⋯H bonds, which meet the above mentioned criteria for the hydrogen bond, were found: H37⋯I1, $d$=3.010 Å, $\alpha$=134.89° and H20⋯I4–Sn, $d$=3.144 Å, $\alpha$=106.74°. In Table 1 and Figure 2c also I⋯H interactions with slightly larger distances are listed. These hydrogen bonds with halogen acceptors obviously contribute to crystal formation, but as we explained further, I⋯H bonds have also an impact on the electronic properties of this material. Also π–π interactions are present in this structure (Figure 2b). Angles between stacked rings seem to be influenced by steric hindrances. Distances between centroids of phenyl rings arranged in parallel (offset) or perpendicular fashion are 4.218 and 5.057Å, respectively. Listed close contacts are responsible for formation of a zigzag pattern present in the crystal structure (Figure 2d). The last interesting factor that may influence electronic properties of this material is the distance between the two closest iodine atoms (Figure 2e). In contrast to [SnI$_4${(C$_6$H$_5$)$_2$SO}$_2$] and [SnI$_4$(C$_5$H$_5$NO)$_2$] complex, the distance I⋯I in [SnI$_4${(C$_6$H$_5$)$_3$PO}$_2$] is quite large, namely 6.015 Å.

*Crystal structure of the [SnI₄{(C₆H₅)₂SO}₂] complex*

The [SnI$_4${(C$_6$H$_5$)$_2$SO}$_2$] complex crystalizes in the orthorhombic space group Pna2$_1$. The Sn$^{4+}$ cation coordinates four iodide anions and two oxygen atoms from diphenyl sulfoxide ligands (Figure 3a). Sn–O1 and Sn–O2 lengths are 2.167 and 2.207 Å, respectively, with O1–Sn1–O2 angle of 86.17°. In the crystal structure π–π interactions have been found (Figure 3b). Offset parallel stacking with distance between centroids 4.068 Å shows slight slope (11.7°). Second π interaction is formed by two almost perpendicular phenyl rings (4.856 Å. 87.5°). As mentioned above, the I⋯H bonding seems to be very important in this material. One strong intermolecular hydrogen bond was found (H32⋯I4, $d$=3.105 Å, $\alpha$ =91.92°) but in the short distance (less than the sum of van der Walls radii + 0.1 Å). Other two I⋯H interactions are present (Figure 3c, Table 1). Interestingly, a close S1⋯I2 distance, 3.651 Å, points at a halogen bond. Crystal packing of complex [SnI$_4${(C$_6$H$_5$)$_2$SO}$_2$] is presented in Figure 3d. The distance between iodine atoms is 4.033 Å (Figure 3e).

*Crystal structure of the [SnI₄(C₅H₅NO)₂] complex.*

The [SnI$_4$(C$_5$H$_5$NO)$_2$] complex (Figure 4a) crystalizes in the triclinic space group P-1. It consist of Sn$^{4+}$ cation with coordinated four iodide anions and two pyridine *N*-oxides where the Sn1–O1 distance is 2.162, Sn1–O2 is 2.168 Å and the O1–Sn1–O2 angle of 82.79°. The crystal packing is presented in Figure 4d. The pyridine rings of neighbouring molecules lay in the same planes. The presence of N$^+$O$^-$ group induces formation of intramolecular hydrogen bond O2⋯H6; 2.686 Å. The coordination of strongly polarised ligand is probably responsible for deviations in electrochemical behaviour of this compound in solution. As in previous structures, hydrogen bonds with halogen acceptor, suspected to contribute strongly to the band structure, are formed: H6⋯I4,

$d$=3.122 Å, $\alpha$=107.38° and H9···I3, $d$=3.085 Å, $\alpha$=151.99° and H10···I3, $d$=3.180 Å, $\alpha$=93.08°. The arrangement of iodide–rings contacts is presented in Figure 4c. In this structure also an intermolecular halogen bond between I1 atom and O2 atom is observed. The π interaction is represented by an offset parallel arrangement of two pyridine rings. The type I halogen contact was found with the 3.837 Å I···I distance (Figure 4e).[32, 33] This and other contact distances are listed in Table 1.

## *$^1$H NMR spectroscopy*

$^1$H NMR spectra of ligands and complexes dissolved in acetonitrile prove that SnI$_4$L$_2$ complexes are present in the solution. Spectra of [SnI$_4$(C$_5$H$_5$NO)$_2$] are strongly influenced by coordination of tin tetraiodide, what reflects in a shift of the pyridine ring signal, even by 0.9 ppm. Moreover, the impact of electro-withdrawing substituent is noticeable: the triplet belonging to the *meta* position in the ring exhibits the biggest shift in the complex (by ca. 0.9 ppm) while signals from protons in *ortho* and *para* positions show a smaller shift, by ca. 0.6 ppm. In [SnI$_4${(C$_6$H$_5$)$_3$PO}$_2$] and [SnI$_4${(C$_6$H$_5$)$_2$SO}$_2$] the shift is smaller (0.04 and 0.01 ppm, respectively Figure 5) The shift towards a lower field means that the electron density is slightly withdrawn from rings. Neat charge transfer values obtained from Charge Transfer Analysis (*vide infra*) correlate with the shifts observed in NMR spectra (Table 2). The charge transfer for [SnI$_4$(C$_5$H$_5$NO)$_2$] is the biggest, and for [SnI$_4${(C$_6$H$_5$)$_2$SO}$_2$] is the smallest. This explains different values of $^1$H NMR shifts.

## *Density of States (DOS) and resistivity measurements*

DOS distribution analysis is useful for determining the contribution of individual atomic orbitals in conduction and valence bands. DOS distribution (Figure 6) were calculated for clusters made of twelve molecules, with the crystal structure arrangement. The PM7 method (implemented within the MOPAC software) including the LCAO-MO approximation was used to calculate the contribution of atomic orbitals to molecular orbitals while AOMix software was used to calculate the DOS and PDOS distributions. A qualitative analysis of DOS distribution reveals that valence bands of studied materials consist of iodide orbitals, while conduction bands are composed of carbon orbitals. This indicates the importance of short distances between iodine atoms of one molecule and conjugated rings of another one (*vide infra*). These interactions, listed in Table 1, provide a significant orbital overlap and hence an efficient electron transfer upon excitation. After excitation, electron is transferred from iodine atom to aromatic rings. Subsequently, it can travel through the crystal structure due to the π–stacked assembly of the rings. One can notice in Figure 6 that the apperance of valence band of $[SnI_4(C_5H_5NO)_2]$ differs from the appearance of the other two complexes: the PDOS distribution of iodine-related states at the top of the valence band is narrower and more intense. This is the results of a short contact between iodine atoms in the crystal structure. In the $[SnI_4(C_5H_5NO)_2]$ complex the I2–I2 distance is only 3.837 Å (Table 1), and the van der Waals spheres of these iodide atoms intersect. In the $[SnI_4\{(C_6H_5)_3PO\}_2]$ complex and $[SnI_4\{(C_6H_5)_2SO\}_2]$ complex the distances are larger and the van der Waals spheres of the closest iodide atoms do not cross, what suggests a much smaller influence of these interactions on the DOS distribution.

Similar results were found in a hybrid $C_7H_7PbI_3$ inorganic-organic material,[10] the valence and conduction bands consist of iodine and carbon orbitals, respectively. Although $C_7H_7PbI_3$ exhibits a strong organic-inorganic coupling, this compound has an ionic structure (organic cation and inorganic anion). Measured with four point method the values of the sheet resistance of thin layers of studied complexes are very large with 1.3 MΩ/□, 0.4 MΩ/□ and 21 MΩ/□ for [SnI$_4${(C$_6$H$_5$)$_3$PO}$_2$], [SnI$_4${(C$_6$H$_5$)$_2$SO}$_2$] and [SnI$_4$(C$_5$H$_5$NO)$_2$] polycrystalline thin films of 35, 63 and 45 nm thicknesses, respectively. These values correspond to specific resitivities of 0.46, 0.24 and 9.45 Ωm at room temperature, which are not worse than those for lead iodide perovskites.[34]

*UV–Vis absorption spectra*

Prepared complexes are soluble in most organic solvents. UV-Vis absorption spectra of all studied complexes (including SnI$_4$) in acetonitrile are very similar (Figure 7a). The only difference is observed for a slightly shifted spectrum of [SnI$_4$(C$_5$H$_5$NO)$_2$]. It is worth noticing, that [1]H NMR spectra in acetonitrile confirm formation of the complexes in solution, therefore the presented UV-Vis absorption spectra are neither the results of dissociation of organic ligands from SnI$_4$ nor solvation of the complexes. TDDFT calculations were applied to explain this observation. Results suggest that the absorption bands of SnI$_4$ in acetonitrile consist of transitions from HOMO (H-1, H-2, H-3, H-4) localized on iodine *p* orbitals to LUMO which is an Sn–I antibonding orbital. Calculations show also that transition between the same orbitals dominates in absorption spectra of [SnI$_4${(C$_6$H$_5$)$_3$PO}$_2$], [SnI$_4${(C$_6$H$_5$)$_2$SO}$_2$] and [SnI$_4$(C$_5$H$_5$NO)$_2$]. Moreover, HOMO and LUMO of isolated complexes in acetonitrile are localized on inorganic fragments (see Figs.S1-S3, supporting information), making all spectra very similar (Figure 7a). Therefore organic ligands have almost no influence on the absorption spectra of these complexes studied in solutions.

Solid state absorption spectra are strikingly different from absorption spectra in solution. Diffuse reflectance spectroscopy allowed to observe new bands in the red part of the spectrum. Owing to their relatively high intensity they can be assigned to a charge transfer character. When organic ligand is coordinated to tin(IV) iodide, a new, very broad absorption band appears, while two characteristic maxima (289 and 354 nm) disappear (Figure 7b). The difference between solid and solution absorption spectra is a result of the presence of intermolecular interactions between complex molecules, especially via the I⋯HC contacts. This interaction is too weak to reveal itself in the UV-Vis absorption spectra in solution, but in the solid crystal structure numerous I⋯HC contacts were encountered (Table 1). Under the influence of the I⋯HC interactions, the electronic structure of complex changed: maxima (289 and 354 nm) vanished, whereas a new broad band extending to the visible range (ca. 300-700 nm) appeared. This new spectral feature may be attributed to the charge transfer process involving iodide ligands (donor) and aromatic rings (acceptor) belonging to adjacent molecules. This assumption is supported by the DOS distribution, where the conduction band is made of carbon atom orbitals and the valence band of the iodine atom orbitals. Therefore, for the isolated molecule of the complex without any I⋯HC interactions, HOMO–LUMO transitions appear on the inorganic ($SnI_4$) moiety. Conversely, spectral properties of the solid phase are determined by the intermolecular interactions within the lattice. The main structural features involved are close I⋯HC contacts.

In order to better explain how the presence of neighboring molecules influences the localization of the frontier orbitals, DFT and TDDFT calculations were performed on molecular dimers with initial geometries extracted from the crystal structures. This computational approach is commonly used in determination of electrical/optical properties of molecular semiconductors.[35-37] More accurate calculations for solids, where three dimensional clusters are made of twelve molecules, are present in the DOS section (*vide supra*). The significance of intermolecular interactions of adjacent molecules appear in the solid state arrangement. As the HOMO of single molecule in solution and dimer in crystal arrangement are alike, the important changes appear in LUMO: for molecule of the complex in solution, HOMO is composed of *p* orbitals of iodide, while LUMO shows the I–Sn antibonding character. The arrangement of atoms in the crystal phase and the presence of neighboring molecule results in HOMO still localized mostly on iodine atoms (Figure 8a, c, and e) of one molecule within a dimer, whereas LUMO possesses a significant contribution of phenyl (or pyridine) rings of the second molecule of the dimer (Figure 8 b, d, and f). These iodine–aromatic ring are characterized by a smaller energy and appear in the visible part of absorption spectra. It is consistent with the results of semiempirical calculations where conduction and valence bands are made of carbon and iodine atom orbitals, respectively.

Diffuse reflectance spectroscopy was employed to determine the band gap energies. By plotting $(F^{KM}*h\nu)^{1/n}$ vs $h\nu$ (Tauc plot) the optical band gap can be obtained. In the case of typical inorganic semiconductors $n = ½$ for direct and $n = 2$ for indirect allowed transitions. In organic and molecular semiconductors, in which a weak van der Waals interactions play an important role in band formation, the $n$ factor is often equal to unity (as used in Figure 9). The analysis of crystallographic structures of the compounds reveals the presence of those weak interactions between individual molecules in the crystals of studied materials. Therefore it seems reasonable to use $n=1$ for the studied compounds.[38] Optical band gaps and calculated by TDDFT lowest transitions (for isolated, optimized molecule in vacuum), are collected in Table 3. Coordination of organic ligands causes a reduction of optical band gap energy by about 1 eV. The lowest transition calculated by TDDFT reflects the excitation of a molecule. Changes in calculated transition energy values (for allowed transitions) show the same trend as changes in experimentally obtained band gap energies.

*Charge Decomposition Analysis*

Bonding in the complex is considered as an interaction of fragment molecular orbitals of two closed-shell fragments. Charge decomposition analysis (CDA) distinguishes three interactions between the metal center and ligands: (i) mixing of occupied orbitals of ligand with unoccupied orbitals of metal – the electron donation, charge transfer from ligand to metal (LMCT), (ii) mixing of unoccupied orbitals of ligand with occupied orbitals of metal – the electron back donation, charge transfer from metal to ligand (MLCT), (iii) repulsive interaction of occupied orbitals of ligand and metal – repulsive polarization.[39] When $SnI_4$ (**1**) and organic ligand (**2**) combine to form a complex, two types of electronic interactions appears: polarization (PL) of one fragment in the presence of another and a charge transfer (CT) from a donor to an acceptor. Electronic polarizations do not move electron density from the donor fragment to acceptor, but cause the electron redistribution within the fragment, in contrast to the CT process.[40] Using quantum-chemical calculations at the DFT level of theory and charge decomposition analysis, the CT and PL interactions were separated (Figure 10). The CT(2→1)-CT(1→2) difference confirms the supposition that $SnI_4$ is an acceptor of electron density and the organic ligand is a donor in the complexes, but more striking are the differences in ΔCT between these three complexes. $[SnI_4(C_5H_5NO)_2]$ stands out with the highest ΔCT and hence, the biggest neat charge transfer value. It means that in $[SnI_4(C_5H_5NO)_2]$ complex bonding between tin and oxygen is the strongest. The higher ΔCT value influences the properties of the complex both in solution and in the solid phase. It correlates with the biggest changes in $^1$H NMR spectra for this complex. Furthermore, it is related with the lowest specific conductivity due to the depletion of π-electron density. ΔPL values for $[SnI_4\{(C_6H_5)_3PO\}_2]$ and $[SnI_4\{(C_6H_5)_2SO\}_2]$ are basically the same, while for $[SnI_4(C_5H_5NO)_2]$ Δp is the smallest which indicates the smallest repulsion between pyridine *n*-oxide and tin tetraiodide.

*Surface photovoltage (SPV) measurements*

Surface photovoltage measurements were used to determine the majority current type. It is a widely used noncontact method, based on changes of surface and near surface potential distribution under illumination. Band bending (upward for *n*-type, downward for *p*-type semiconductors) is a result of many factors, including surface dipole, termination of crystal structure, doping, surface states, crystal structure defects and many others. Under illumination of the semiconductor surface, minority current carriers travel to the surface causing flattening of bands. The measurements of ΔCPD allow to determine the type of conductivity. In general:

$$e\Delta CPD = -e\Delta SPV$$

where *e* is the electric charge, ΔCPD is the contact potential difference, SPV is the surface photovoltage.[41] Differential surface photovoltage spectra of examined compounds (Figure 11) point at *n*-type of tin tetraiodide complexes (negative ΔCPD) with a small excess of majority current curriers (a small absolute value of ΔCPD). *N*-type conductivity might be caused by I⁻ vacancies, associated with a partial reduction of Sn(IV) centers. The *n*-type conductivity may be caused by a partial reduction of Sn(IV) centers (and hence iodide depletion).[42, 43] Iodide vacancies in similarly composed perovskites were described in literature as the primary source of charge carriers,[44] this statement also supports observation of iodine sublimation from long-time stored materials. The spectral range of SPV response (UV and Vis) is consistent with solid state absorption spectra.

*Work function (WF) and Fermi level (FL)*

Measurement of the work function (WF) allows to determine one of the most basic parameter in semiconductors studies – the potential of Fermi level (FL). Work function is the energy required to remove one electron from the Fermi level. Using a Kelvin probe technique WF and FL have been determined (Table 4). Tin(IV) iodide in the studied complexes acts as an acceptor of electron density while the ligand acts as a donor. Upon coordination of organic ligand to $SnI_4$ WF decreases, due to a higher electron density on $SnI_4$ fragment of the complex. Energy of frontier orbitals were obtained from TDDFT calculations made for molecules with vacuum-optimized geometries. HOMO and LUMO of complexes are shifted with respect to tin tetraiodide corresponding orbitals, which is consistent with decreasing values of work function observed upon coordination of organic ligands to $SnI_4$ (Figure 12). Due to a significant discrepancy between predicted and observed electron affinities (LUMO energy), which is a well-known problem of the DFT method,[45] the value of original LUMO energies were corrected by adding to HOMO energies the value of the lowest allowed transitions (Table 3) to obtain a more reasonable energy diagram. This correction is often used to calculate LUMO energy levels.[46] Theoretical values of Fermi level energies were estimated on the basis of DFT calculations and Janak's theorem as follows:[47]

$$E_F = \frac{E_{HOMO} + E_{LUMO}}{2}$$

where $E_F$ is Fermi level energy, $E_{HOMO}$ and $E_{LUMO}$ are energies of HOMO and LUMO orbitals. The results of these calculations show that theoretical and measured Fermi levels of complexes are consistent and localized around the middle of energy gap. For SnI$_4$ the measured Fermi level is located close to the LUMO level and significantly differs from the theoretical Fermi level energy, which indicates more defects in SnI$_4$.

**Conclusions and outlook**

The current study can be summarized in four points indicating the main findings regarding the chemistry and physics of classical tin(IV) complexes.

The analysis of crystal structures of the synthesized tin(IV) tetraiodide complexes shows the presence of π interactions along with hydrogen bonds (H⋯I). Due to the presence of only weak intermolecular interactions, SnI$_4$ complexes are unique among the hybrid materials. Moreover, the composition of band structure highlight the importance of weak interactions in solid state: it causes the formation of the valence band mainly from the orbitals belonging to the inorganic core and the conduction band from the organic-ligand-centered orbitals. Also short I⋯I contacts were found to contribute significantly to the band structure of [SnI$_4$(C$_5$H$_5$NO)$_2$]. Their distances are smaller than in the SnI$_4$ crystal structure (4.206 Å).

Solid state and solution absorption spectra are completely different for the studied complexes. A new, very broad charge transfer band appears in solid state spectra. TDDFT analysis of complexes in solution and in solid shows that the transitions in visible part of spectra (solid absorption) are between organic and inorganic fragments of neighboring molecules. In solution, due to a lack of weak intermolecular interaction, both HOMO and LUMO are localized at $SnI_4$.

Charge Decomposition Analysis explains differences between measured properties of complexes under conditions, where intermolecular interactions are absent, *i.e.* in acetonitrile solution. Therefore proton signals in $^1$H NMR spectra exhibit the smallest shift for the complex with the smallest charge transfer ($[SnI_4\{(C_6H_5)_2SO\}_2]$) and the biggest shift for the complex with the biggest charge transfer ($[SnI_4(C_5H_5NO)_2]$).

Work function measurements show how coordination of organic ligand decreases the Fermi level potential. Measured Fermi levels are localized close to the centre of calculated energy gaps of complexes, which is characteristic for molecular semiconductors. Surface photovoltage measurements revealed that these complexes are *n*-type semiconductors with small excess of majority charge current carriers.

The detailed studies presented above indicate the crucial role of intermolecular interactions in design of novel semiconducting materials. Structural and physicochemical analysis helped to reveal a new and exciting family of molecular semiconductors based on main group elements. The main features of these materials are good solubility in common solvents (what enables e.g. application of printing technologies to produce $SnI_4L_2$–based devices) and possible tunability of properties. Tin iodide offers a few and easily accessible platform for numerous semiconducting materials, as the variety of ligands it can accept is almost unlimited.

**Supporting information available:** HOMO and LUMO of isolated complexes in acetonitrile (Figs.S1-S3) and cif files containing crystallographic data of studied compounds obtained by single crystal diffraction. This material is available free of charge via the Internet at http://pubs.acs.org.

**Acknowledgements**

Authors acknowledge the support of Polish Ministry of Science and Higher Education (grant Ideas Plus No. IDP2012000362), National Science Centre (Poland) within the MAESTRO project (grant No. UMO-2015/18/A/ST4/00058) and European Union (RECORD–IT within the H2020-FETOPEN-1-2014 project, grant No. 664786).

**Bibliography**

1. Klauk, H., *Organic electronics: materials, manufacturing and applications*. Wiley-VCH: Weinheim, 2006.
2. Brütting, W., *Physics of organic semiconductors*. Wiley-VCH: Weinheim, 2005.
3. Schwoerer, M.; Wolf, H. C., *Organic molecular solids*. Wiley-VCH: Weinheim, 2007.
4. Kagan, C. R.; Mitzi, D. B., *Science* **1999,** *286*, 945-947.
5. Knutson, J. L.; Martin, J. D.; Mitzi, D. B., *Inorganic Chemistry* **2005,** *44*, 4699-4705.
6. Mitzi, D. B., *Inorganic Chemistry* **2000,** *39*, 6107-6113.
7. Yu, T. L.; Zhang, L.; Shen, J. J.; Fu, Y. B.; Fu, Y. L., *Dalton Transactions* **2014,** *43*, 13115-13121.
8. She, Y. J.; Zhao, S. P.; Tian, Z. F.; Ren, X. M., *Inorganic Chemistry Communications* **2014,** *46*, 29-32.
9. Zhu, X. H.; Mercier, N.; Frere, P.; Blanchard, P.; Roncali, J.; Allain, M.; Pasquier, C.; Riou, A., *Inorganic Chemistry* **2003,** *42*, 5330-5339.
10. Maughan, A. E.; Kurzman, J. A.; Neilson, J. R., *Inorganic Chemistry* **2015,** *54*, 370-378.
11. Dickinson, R. G., *Journal of the American Chemical Society* **1923,** *45*, 958-962.
12. Jatsenko, A. V.; Medvedev, S. V.; Paseshnitchenko, K. A.; Aslanov, L. A., *Journal of Organometallic Chemistry* **1985,** *284*, 181-188.
13. Tudela, D.; J. Sanchez-Herencia, A.; Diaz, M.; Fernandez-Ruiz, R.; Menendez, N.; D. Tornero, J., *Dalton Transactions* **1999**, 4019-4023.
14. Tursina, A. I., Aslanov, L. A., Chernyshev, V. V., Medvedev, S. V. & Yatsenko, A. V, *Koordinatsionnaya Khimiya* **1986,** *12*, 420 - 424.
15. *CrysAlis PRO. Oxford Diffraction Ltd Yarnton, England* **2010**.
16. Altomare, A.; Burla, M. C.; Camalli, M.; Cascarano, G. L.; Giacovazzo, C.; Guagliardi, A.; Moliterni, A. G. G.; Polidori, G.; Spagna, R., *Journal of Applied Crystallography* **1999,** *32*, 115-119.
17. Sheldrick, G. M., *Acta Crystallographica* **2008,** *A64*, 112-122.

18. Macrae, C. F.; Edgington, P. R.; McCabe, P.; Pidcock, E.; Shields, G. P.; Taylor, R.; Towler, M.; Van de Streek, J., *Journal of Applied Crystallography* **2006,** *39*, 453-457.
19. Farrugia, L. J., *Journal of Applied Crystallography* **1999,** *32*, 837-838.
20. Hanwell, M. D.; Curtis, D. E.; Lonie, D. C.; Vandermeersch, T.; Zurek, E.; Hutchison, G. R., *Journal of Cheminformatics* **2012,** *4*, 17.
21. Frisch, M. J.; Trucks, G. W.; Schlegel, H. B., *Gaussian, Inc., Wallingford CT* **2009**.
22. Petersson, G. A.; Al-Laham, M. A., *The Journal of Chemical Physics* **1991,** *94*, 6081-6090.
23. Godbout, N.; Salahub, D. R.; Andzelm, J.; Wimmer, E., *Canadian Journal of Chemistry* **1992,** *70*, 560-571.
24. Kim, K.; Jordan, K. D., *The Journal of Physical Chemistry* **1994,** *98*, 10089-10094.
25. Stephens, P. J.; Devlin, F. J.; Chabalowski, C. F.; Frisch, M. J., *The Journal of Physical Chemistry* **1994,** *98*, 11623-11627.
26. O'Boyle, N. M.; Tenderholt, A. L.; Langner, K. M., *Journal of Computational Chemistry* **2008,** *29*, 839-45.
27. Dennington, R.; Keith, T.; Millam, J., *Semichem Inc., Shawnee Mission, KS,* **2009**.
28. Stewart, J. J. P., *Stewart Computational Chemistry, Colorado Springs CO, USA*.
29. Gorelsky, S. I.; Lever, A. B. P., *Journal of Organometallic Chemistry* **2001,** *635*, 187-196.
30. S. I. Gorelsky AOMix: Program for Molecular Orbital Analysis,version X.X.
31. Brammer, L.; Bruton, E. A.; Sherwood, P., *Crystal Growth & Design* **2001,** *1*, 277-290.
32. Gilday, L. C.; Robinson, S. W.; Barendt, T. A.; Langton, M. J.; Mullaney, B. R.; Beer, P. D., *Chemical Reviews* **2015,** *115*, 7118-7195.
33. Cavallo, G.; Metrangolo, P.; Milani, R.; Pilati, T.; Priimagi, A.; Resnati, G.; Terraneo, G., *Chemical Reviews* **2016,** *116*, 2478-2601.
34. Stoumpos, C. C.; Kanatzidis, M. G., *Accounts of Chemical Research* **2015,** *48*, 2791-2802.
35. Nouri, H.; Cadiou, C.; Lawson-Daku, L. M.; Hauser, A.; Chevreux, S.; Dechamps-Olivier, I.; Lachaud, F.; Ternane, R.; Trabelsi-Ayadi, M.; Chuburu, F.; Lemercier, G., *Dalton Transactions* **2013,** *42*, 12157-12164.
36. Baumeier, B.; Kirkpatrick, J.; Andrienko, D., *Physical Chemistry Chemical Physics* **2010,** *12*, 11103-11113.
37. Brédas, J. L.; Calbert, J. P.; da Silva Filho, D. A.; Cornil, J., *Proceedings of the National Academy of Sciences* **2002,** *99*, 5804-5809.
38. Singh, J., *Advances in amorphous semiconductors / Jai Singh and Koichi Shimakawa*. Taylor & Francis: London, 2003.
39. Frenking, G.; Pidun, U., *Journal of the Chemical Society, Dalton Transactions* **1997**, 1653-1662.
40. Gorelsky, S. I.; Ghosh, S.; Solomon, E. I., *Journal of the American Chemical Society* **2006,** *128*, 278-90.
41. Kronik, L.; Shapira, Y., *Surface photovoltage phenomena: theory, experiment, and applications*. 1999; Vol. 37, p 1-206.
42. Frolova, L. A.; Dremova, N. N.; Troshin, P. A., *Chemical Communications* **2015,** *51*, 14917-14920.
43. Eames, C.; Frost, J. M.; Barnes, P. R. F.; O'Regan, B. C.; Walsh, A.; Islam, M. S., *Nature Communications* **2015,** *6*, 7497.

44. Zheng, F.; Saldana-Greco, D.; Liu, S.; Rappe, A. M., *The Journal of Physical Chemistry Letters* **2015,** *6*, 4862-4872.
45. Zhang, G.; Musgrave, C. B., *Journal of Physical Chemistry A* **2007,** *111*, 1554-1561.
46. Zhang, L.; Shen, W.; He, R.; Liu, X.; Tang, X.; Yang, Y.; Li, M., *Organic Electronics* **2016,** *32*, 134-144.
47. Janak, J. F., *Physical Review B* **1978,** *18*, 7165-7168.

# SEMICONDUCTORS

## INORGANIC

COMPOUND

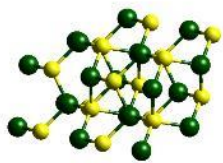

Zinc sulfide

ELEMENTAL

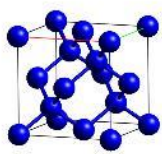

Silicon

## ORGANIC

POLYMERS

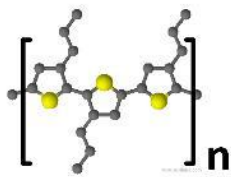

Poly(3-propyl)tiophene

SMALL MOLECULES

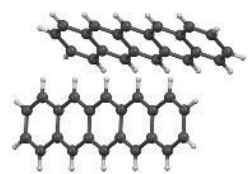

Pentacene

## HYBRID

HELD BY IONIC ATTRACTION

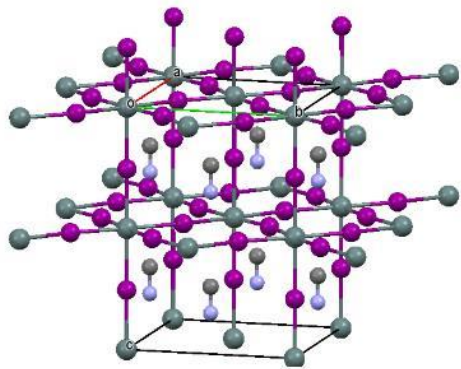

$CH_3NH_3PbI_3$

HELD BY WEAK INTERACTIONS

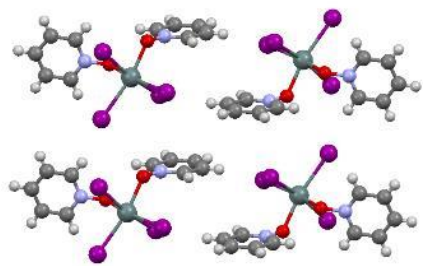

$[SnI_4(C_5H_5NO)_2]$

**Figure 1.** A general classification of semiconductors on the basis of their chemical characteristics.

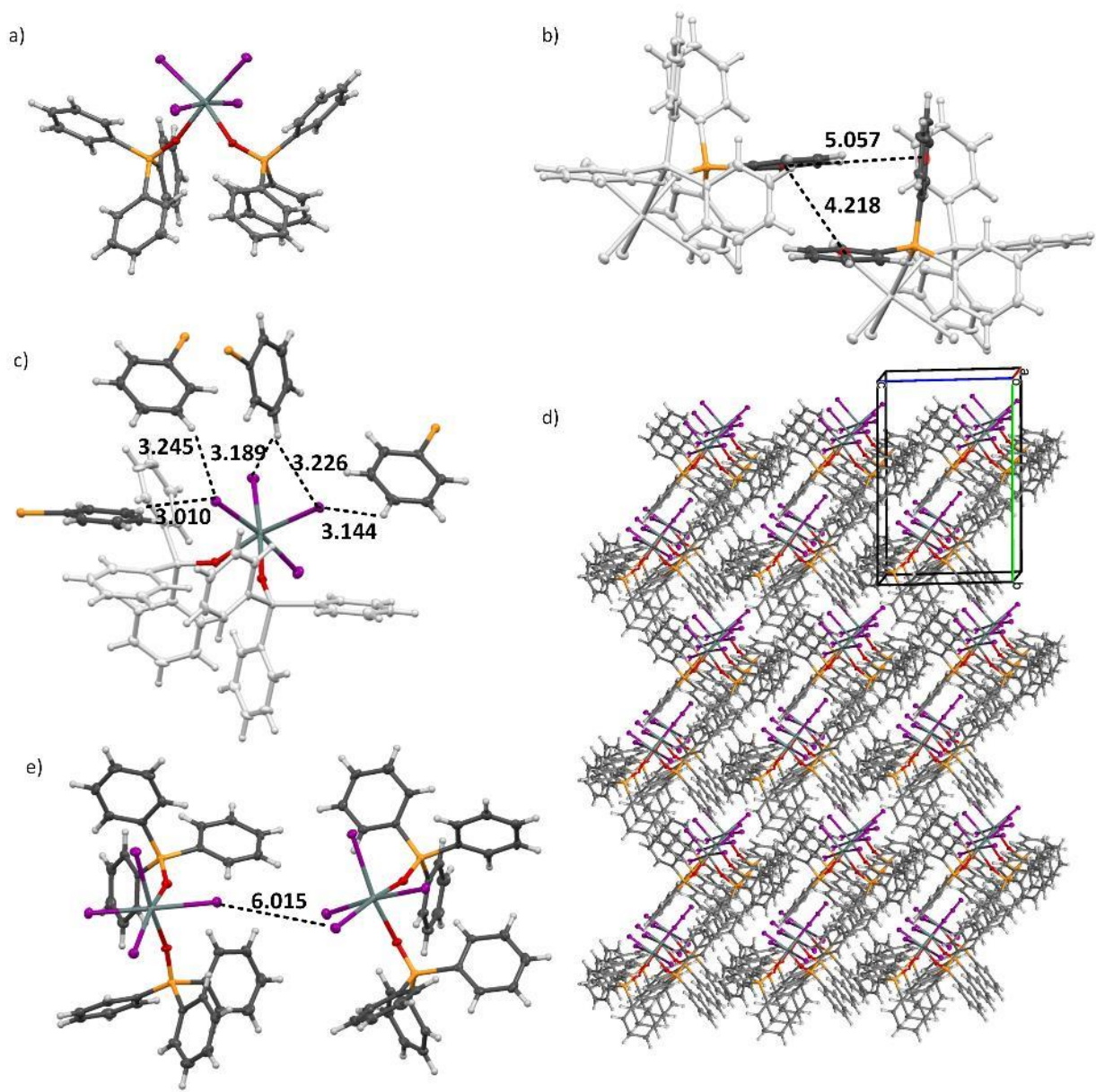

**Figure 2.** Crystal structures of single [SnI$_4${(C$_6$H$_5$)$_3$PO}$_2$] molecule (a), intermolecular stacking of phenyl rings arranged in parallel (offset) or perpendicular fashion (black dashed lines indicate distances between centroids (b), distances (black dashed lines) and aromatic rings arrangement in H···I interaction (only intermolecular aromatic rings are highlighted) (c), packing 3×3×3 and monoclinic cell (d), I···I distance (black dashed line), (e). Ellipsoids represent 50% probability.

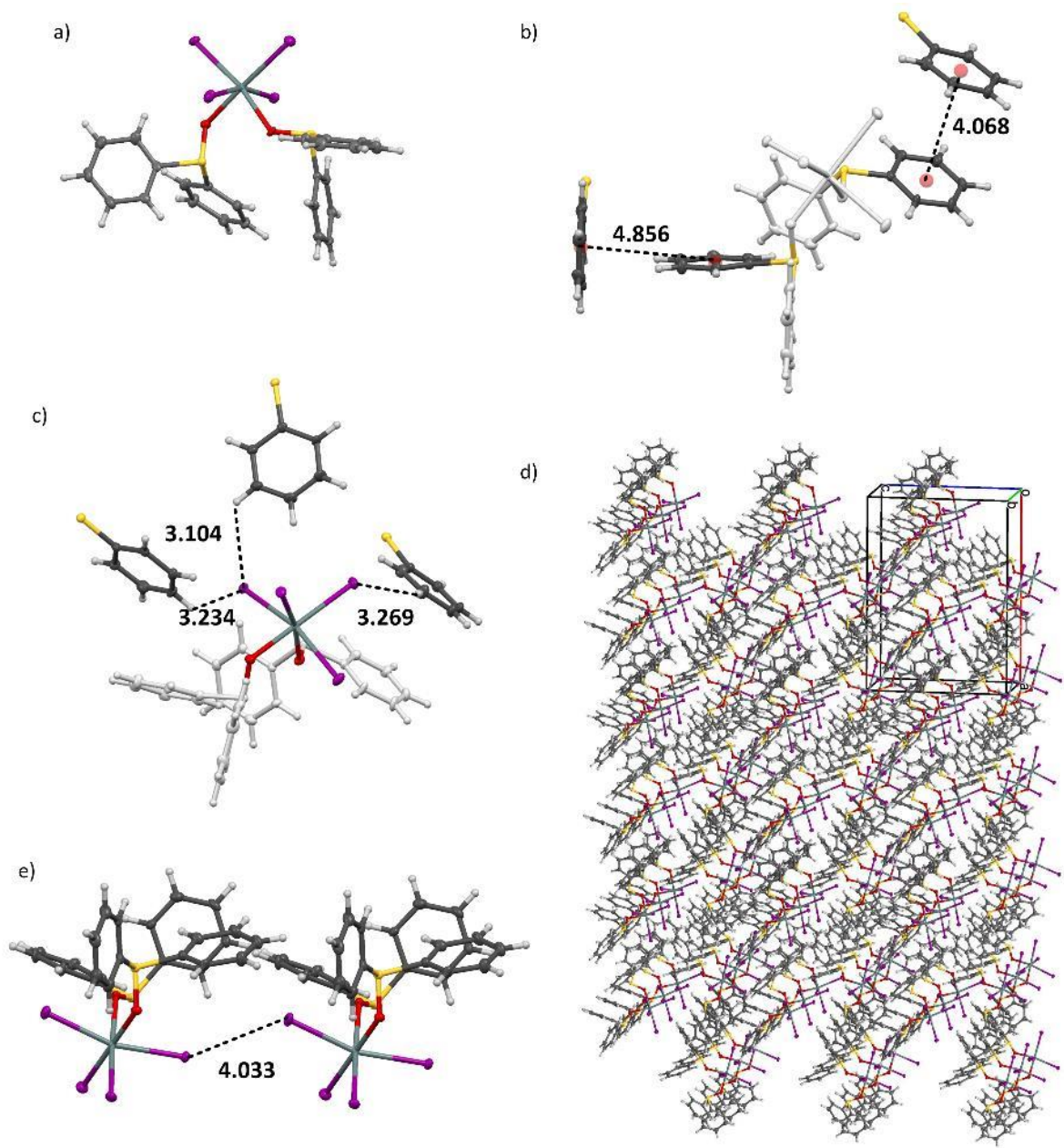

**Figure 3.** Crystal structures of single [SnI$_4${(C$_6$H$_5$)$_2$SO}$_2$] molecule (a), intermolecular stacking of phenyl rings arranged in parallel (offset) or perpendicular fashion (black dashed line indicates distances between centroids (b), distances (black dashed lines) and aromatic rings arrangement in H···I interaction (only intermolecular aromatic rings are highlighted) (c), packing 3×3×3 and orthorhombic cell (d), I···I distance (black dashed lines) (e). Ellipsoids represent 50% probability.

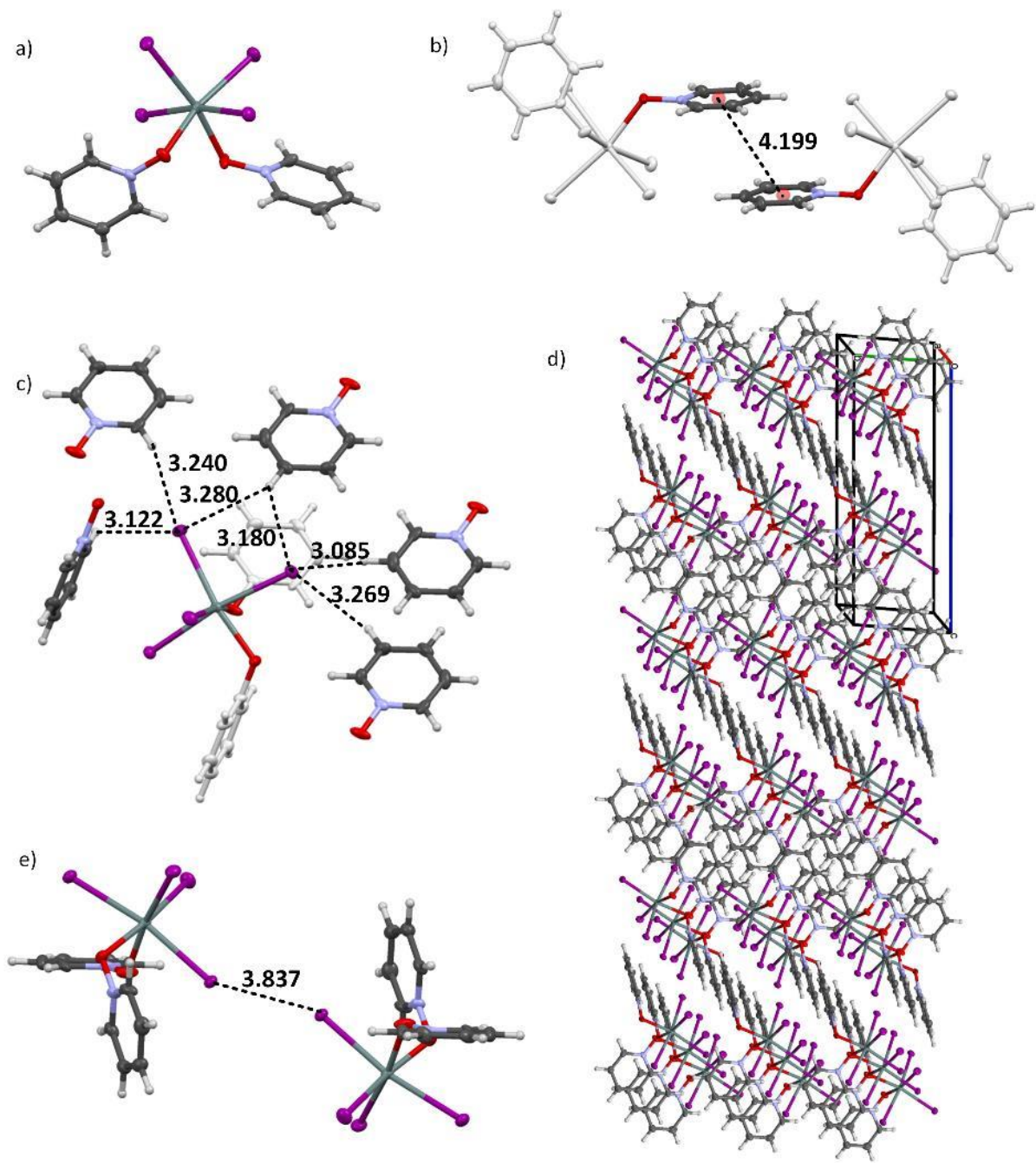

**Figure 4.** Crystal structures of single [SnI$_4$(C$_5$H$_5$NO)$_2$] molecule (a), intermolecular stacking of pyridine rings arranged in parallel (offset), (black dashed line indicates the distance between centroids (b), distances (black dashed lines) and aromatic rings arrangement in H···I interaction (only intermolecular aromatic rings are highlighted) (c), packing 3×3×3 and triclinic cell (d), I···I distance (black dashed line) (e). Ellipsoids represents 50% probability.

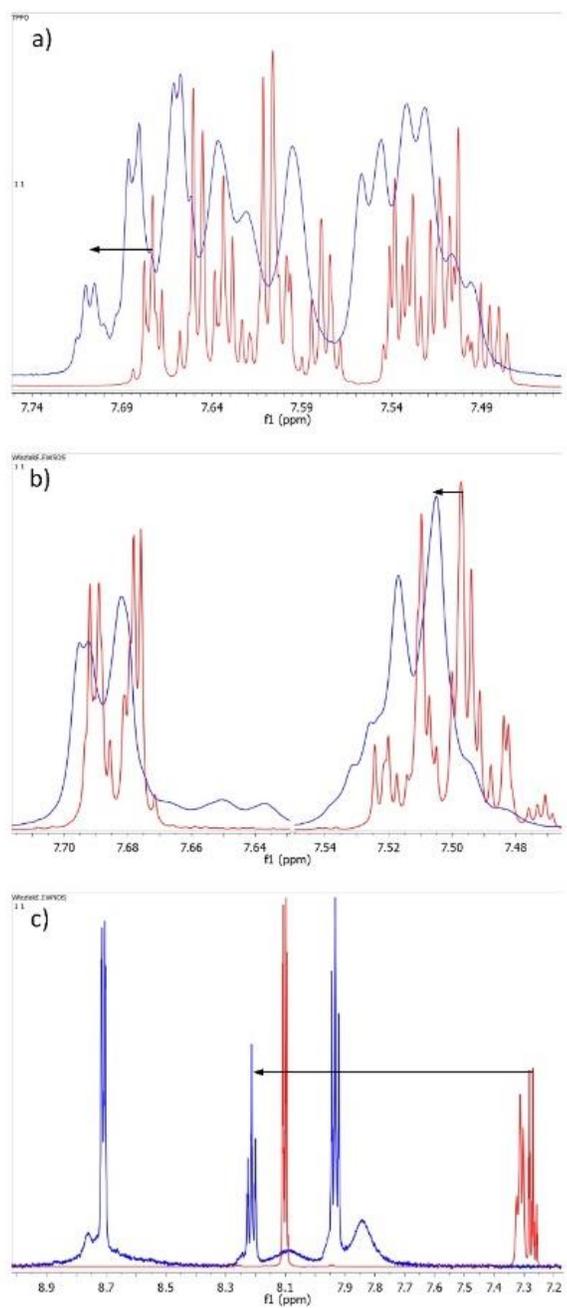

**Figure 5.** $^1$H NMR spectra of $[SnI_4\{(C_6H_5)_3PO\}_2]$ (a), $[SnI_4\{(C_6H_5)_2SO\}_2]$ (b) and $[SnI_4(C_5H_5NO)_2]$ (c) in deuterated acetonitrile. Spectra of the complexes are marked in blue, whereas spectra of free ligands are marked in red.

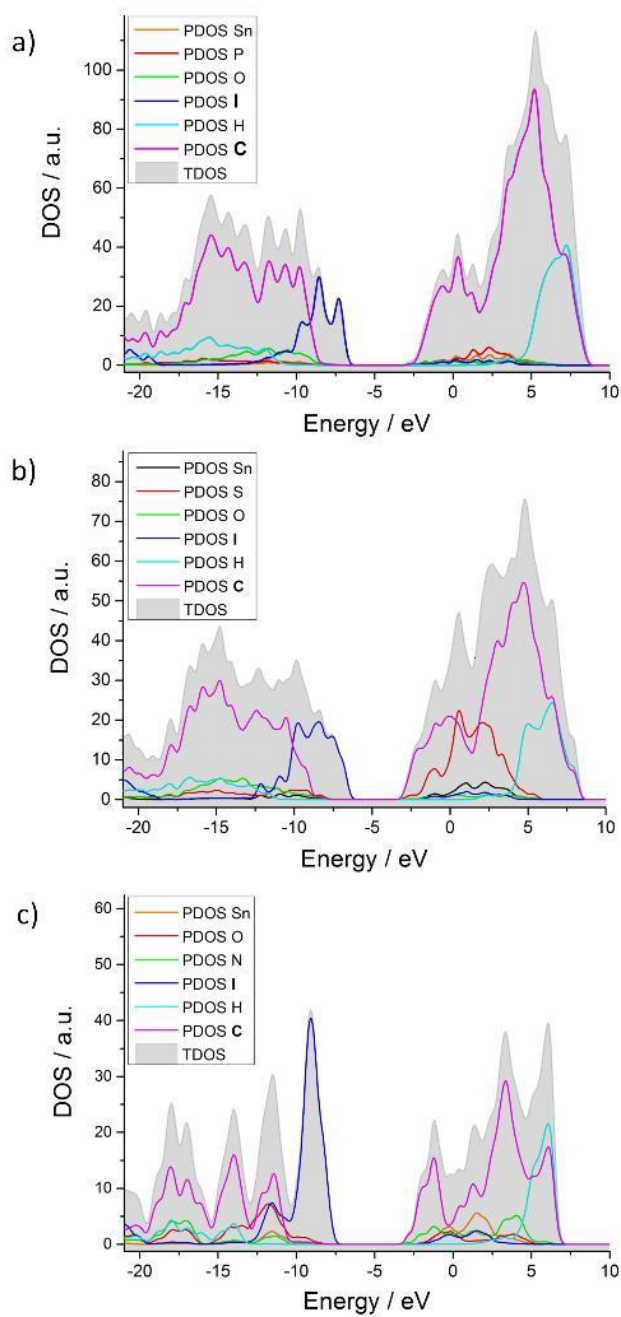

**Figure 6**. Density of states distribution of $[SnI_4\{(C_6H_5)_3PO\}_2]$ (a), $[SnI_4\{(C_6H_5)_2SO\}_2]$ (b) and $[SnI_4(C_5H_5NO)_2]$ (c). Grey area represents the total density of states, colour lines represent partial densities of states.

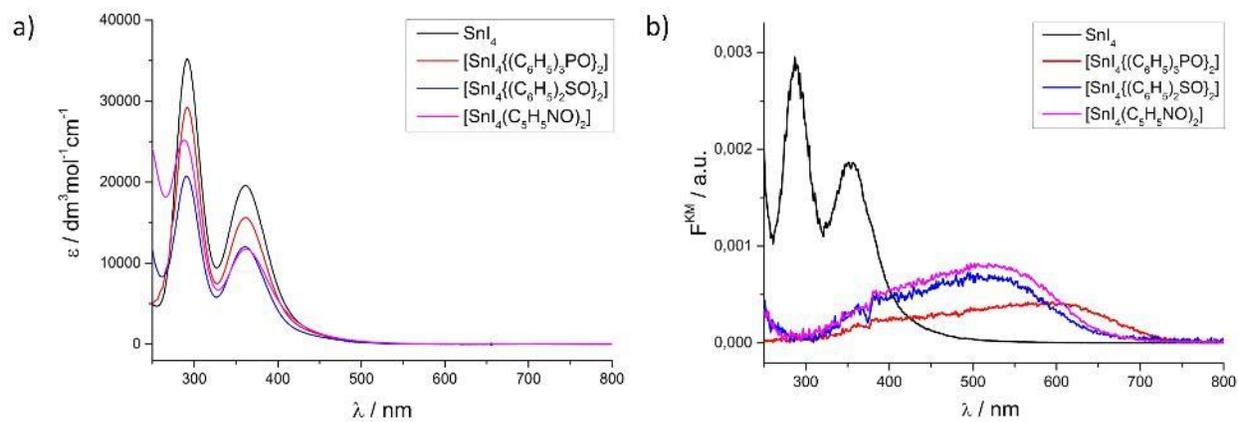

**Figure 7.** UV-Vis absorption spectra of tin tetraiodide complexes in acetonitrile (a) and solid state absorption spectra of tin tetraiodide and complexes (b). $F^{KM}$ is the Kubelka-Munk function.

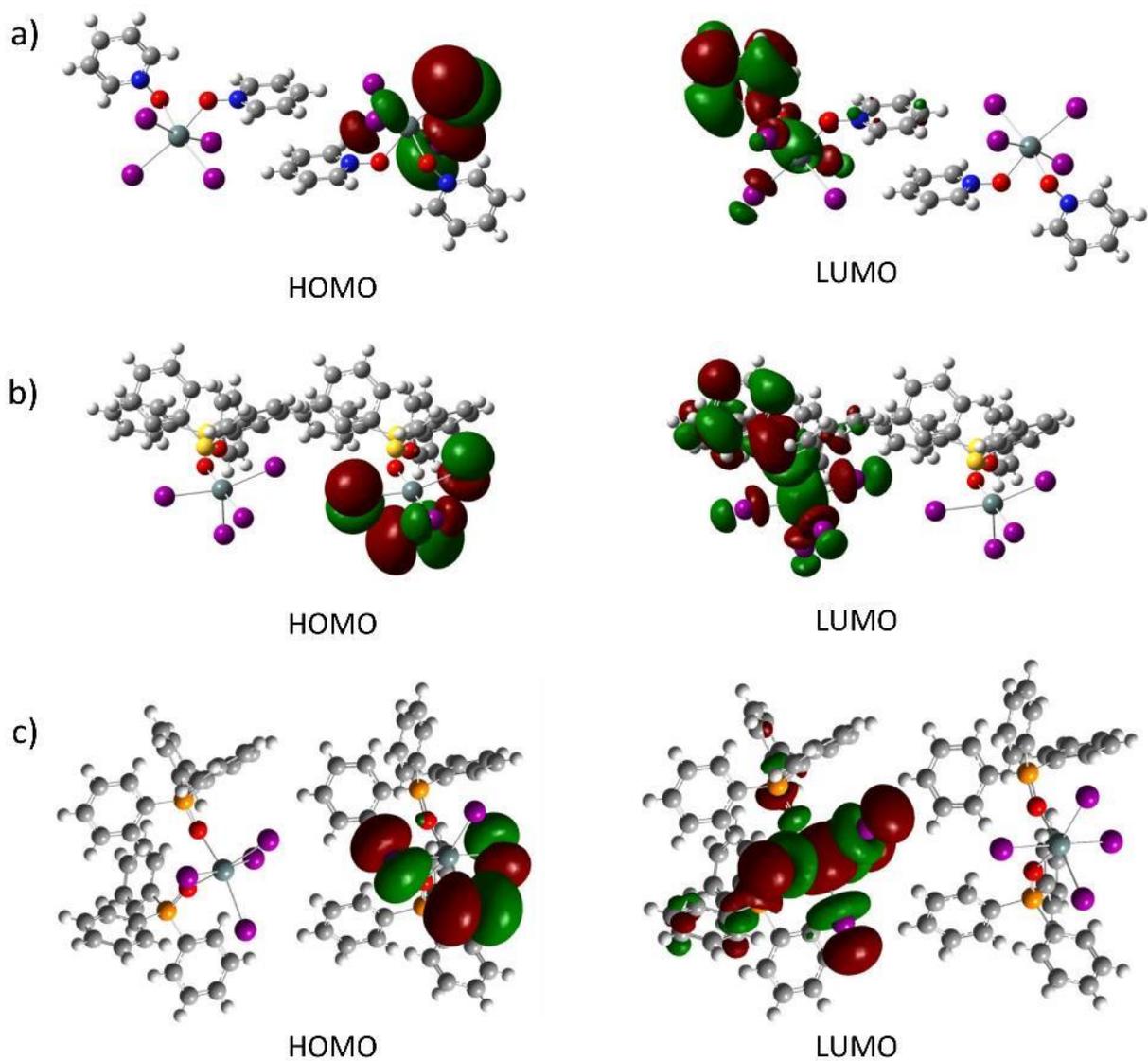

**Figure 8.** Visualisation of frontier molecular orbitals of the dimers of [SnI$_4$(C$_5$H$_5$NO)$_2$] (a), [SnI$_4${(C$_6$H$_5$)$_2$SO}$_2$] (b), and [SnI$_4${(C$_6$H$_5$)$_3$PO}$_2$] (c). HOMO orbitals are localized on *p* iodine orbitals, whereas LUMO orbitals are localized mainly on aromatic carbon rings. HOMO and LUMO are on the opposite sides of the dimer.

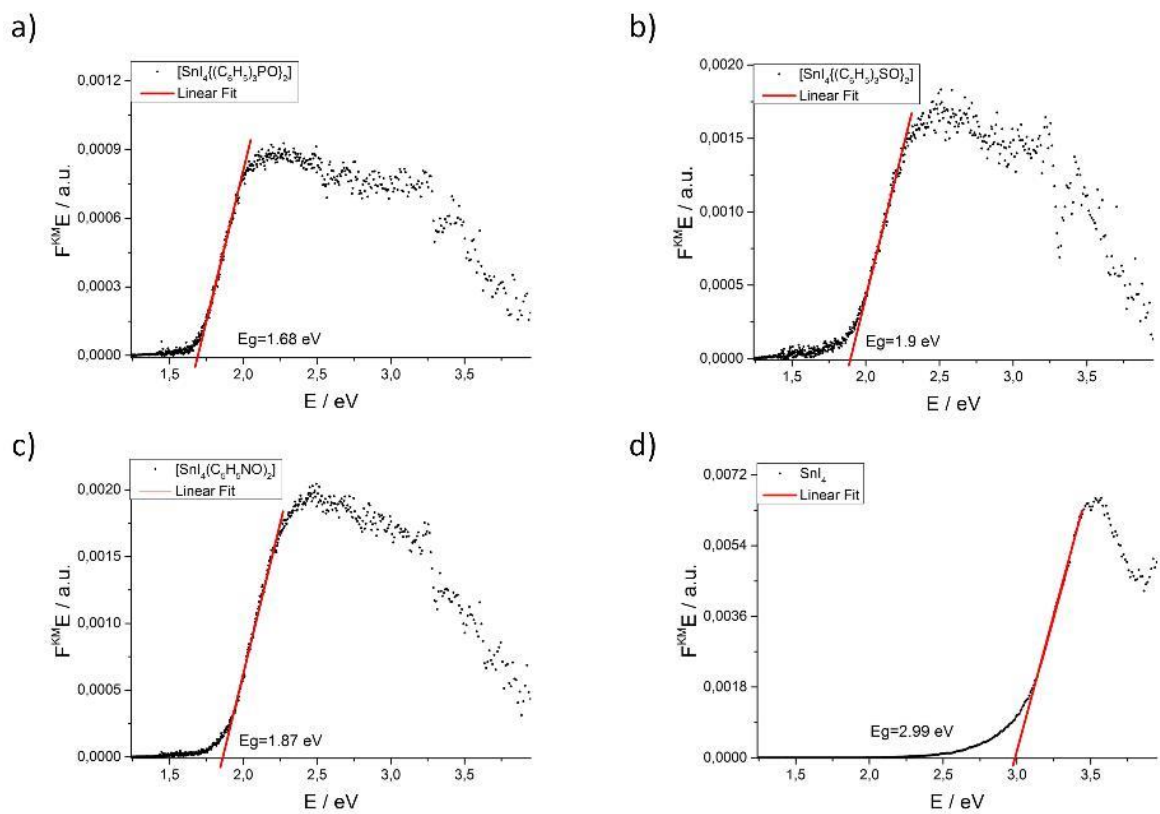

**Figure 9.** Tauc plots for [SnI$_4${(C$_6$H$_5$)$_3$PO}$_2$] (a), [SnI$_4${(C$_6$H$_5$)$_2$SO}$_2$] (b), [SnI$_4$(C$_5$H$_5$NO)$_2$] (c) and SnI$_4$ (d).

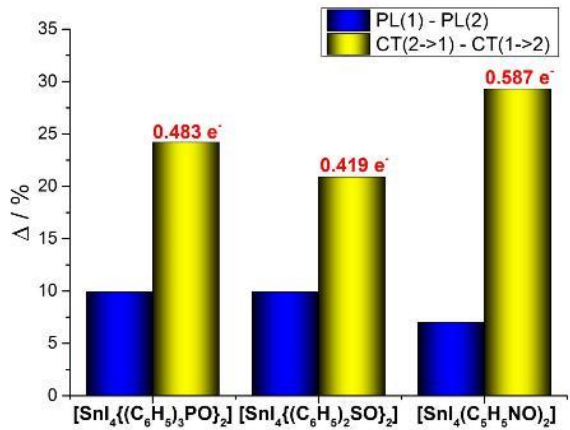

**Figure 10.** Differences between polarizations (ΔPL, blue), charge transfer processes (ΔCT, yellow) and the neat charge transfer (red) in tin(IV) complexes.

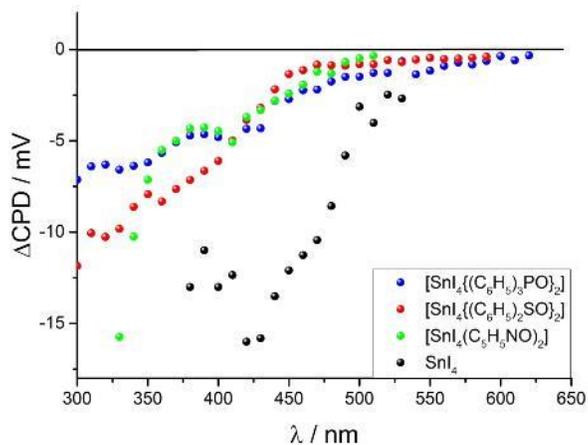

**Figure 11.** Differential surface photovoltage spectra of tin tetraiodide complexes deposited on the surface of ITO foil. Negative ΔCPD indicates the n-type conductivity.

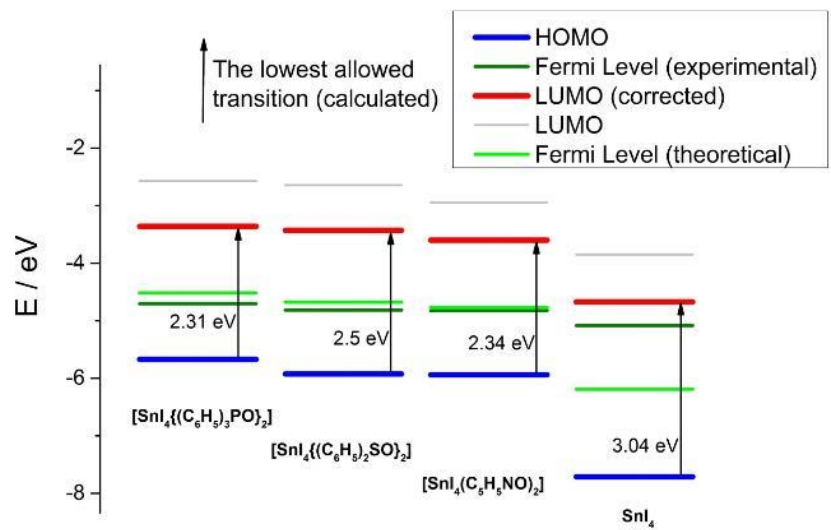

**Figure 12.** Energies of frontier orbitals (HOMO – blue, LUMO – grey) of examined complexes and SnI$_4$. The values of the first allowed transition (next to arrow) were used to correct the LUMO position (red). Fermi levels are depicted in green.

**Table 1.** Intermolecular nonbonded contact distances and angles for phenyl rings and various atom pairs.

| Interaction | [SnI$_4${(C$_6$H$_5$)$_3$PO}$_2$] | [SnI$_4${(C$_6$H$_5$)$_2$SO}$_2$] | [SnI$_4$(C$_5$H$_5$NO)$_2$] |
|---|---|---|---|
| π⋯π[a], ∠π–π[b] | π$_{II}$, 4.218 Å, 4.2°  π$_⊥$, 5.057 Å, 67.4° | π$_{II}$, 4.068 Å, 11.7°  π$_⊥$, 4.856 Å, 87.5° | π$_{II}$, 4.199 Å, 0° |
| A⋯I | - | **S1⋯I2, 3.651 Å** | **O1⋯I1, 3.463 Å** |
| I⋯I | I2⋯I3 6.015 Å | I1⋯I4 4.033 Å | **I2⋯I2 3.837 Å** |
| H⋯I, ∠Sn–I⋯H | **H37⋯I1 3.010 Å**, 134.89°  **H20⋯I4 3.144 Å**, 106.74°  H19⋯I1 3.245 Å  H45⋯I2 3.189 Å  H45⋯I4 3.226 Å | **H32⋯I4 3.105 Å**, 91.92°  H21⋯I3 3.268 Å  H26⋯I4 3.234 Å | **H9⋯I3 3.085 Å**, 151.99°  **H10⋯I3 3.180 Å**, 93.08°  **H6⋯I4 3.122 Å**, 107.38°  H3⋯I2 3.269 Å  H11⋯I3 3.269 Å  H8⋯I4 3.240 Å  H10⋯I4 3.280 Å |

[a] centroid-centroid distance

[b] angle between ring planes

**bolded** – distance shorter than the sum of van der Waals radii

**Table 2.** Neat charge transfer values and chemical shifts.

|  | [SnI$_4${(C$_6$H$_5$)$_3$PO}$_2$] | [SnI$_4${(C$_6$H$_5$)$_2$SO}$_2$] | [SnI$_4$(C$_5$H$_5$NO)$_2$] |
|---|---|---|---|
| Neat charge transfer | 0.483 e$^-$ | 0.419 e$^-$ | 0.587 e$^-$ |
| Chemical shift | 0.04 ppm | 0.01 ppm | 0.9 – 0.6 ppm |

**Table 3.** Optical energy gaps and the lowest calculated transitions.

| Compound | Optical band gap / eV | Calculated transitions / eV |
|---|---|---|
| [SnI$_4${(C$_6$H$_5$)$_3$PO}$_2$] | 1.68 | 2.31 |
| [SnI$_4${(C$_6$H$_5$)$_2$SO}$_2$] | 1.90 | 2.48*, 2.50 |
| [SnI$_4$(C$_5$H$_5$NO)$_2$] | 1.87 | 2.29*, 2.34 |
| SnI$_4$ | 2.99 | 3.04 |

*The first transition with oscillator strength equal zero.

**Table 4.** Work function and Fermi levels of tin tetraiodide complexes.

| Compound | Work Function / eV | Fermi level / mV vs NHE (-4.4+WF) |
|---|---|---|
| [SnI$_4${(C$_6$H$_5$)$_3$PO}$_2$] | 4.704 | 304 |
| [SnI$_4${(C$_6$H$_5$)$_2$SO}$_2$] | 4.815 | 415 |
| [SnI$_4$(C$_5$H$_5$NO)$_2$] | 4.825 | 425 |
| SnI$_4$ | 5.084 | 684 |

TOC FIGURE

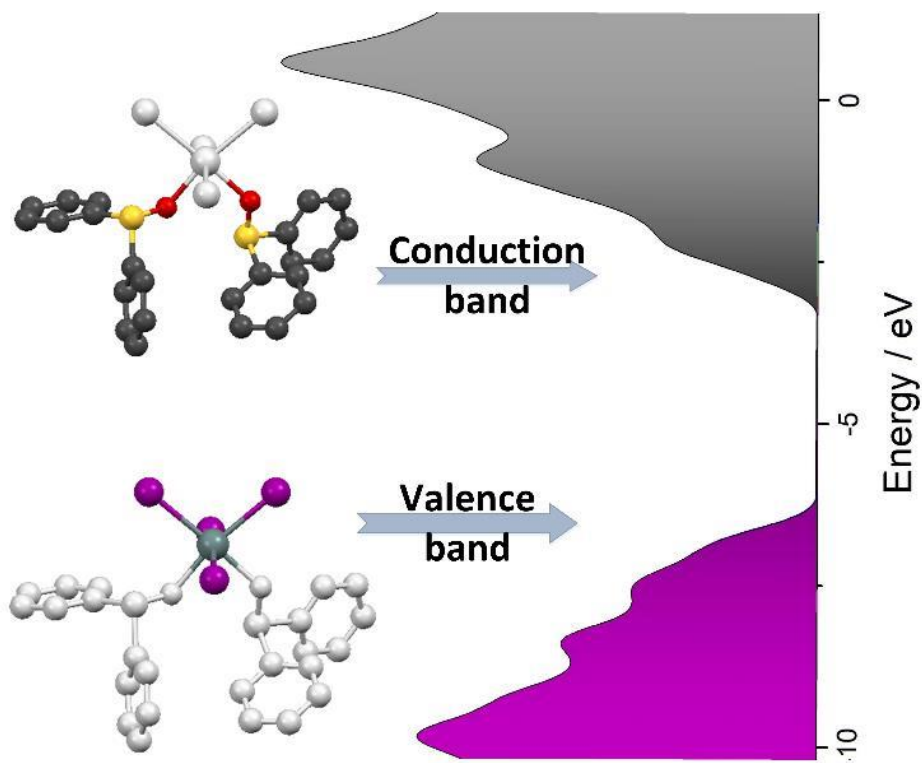

TOC SYNOPSIS

Coordination compounds which combine metal centre, organic and inorganic ligands are a new class of molecular hybrid semiconductors. Experimental and theoretical approaches have been used to support the assumption that in these materials weak intermolecular interactions between inorganic and organic ligands are essential in a band structure formation.